# Next Generation of Phishing Attacks using AI powered Browsers

Akshaya Arun
*Department of Computer Science and Engineering*
*Northumbria University*
London, United Kingdom
a.arun@northumbria.ac.uk

Nasr Abosata
*Department of Computer Science and Engineering*
*Northumbria University*
London, United Kingdom
nasr.abosata@northumbria.ac.

*Abstract*— The increase in the number of phishing demands innovative solutions to safeguard users from phishing attacks. This study explores the development and utilization of a real-time browser extension integrated with machine learning model to improve the detection of phishing websites. The results showed that the model had an accuracy of 98.32%, precision of 98.62%, recall of 97.86%, and an F1-score of 98.24%. When compared to other algorithms like Support Vector Machine, Naïve Bayes, Decision Tree, XGBoost, and K Nearest Neighbor, the Random Forest algorithm stood out for its effectiveness in detecting phishing attacks. The zero-day phishing attack detection testing over a 15-day period revealed the model's capability to identify previously unseen threats and thus achieving an overall accuracy rate of 99.11%. Furthermore, the model showed better performance when compared to conventional security measures like Google Safe Browsing. The model had successfully detected phishing URLs that evaded detection by Google safe browsing. This research shows how using machine learning in real-time browser extensions can defend against phishing attacks. It gives useful information about cybersecurity and helps make the internet safer for everyone.

## I. Introduction

In today's digital landscape, where online interactions are common, cyber risks such as phishing attacks affect users security and privacy. Large phishing campaigns can impact millions of individuals by stealing personal data, installing ransomware and other malware, and getting access to the most sensitive components of a company's networks [1]. Attackers have invented a wide range of phishing strategies to attack different technologies, trends, sectors, and users [14]. Most phishing victims either lack awareness of how they were targeted or are emotionally manipulated into falling for the scam [11]. The three commonly used methods are the text messaging, internet, and phone calls [2]. Phishing can occur through Internet, such as email, social networks, websites, and Wi-Fi [2]. Phishing detection technologies have low detection accuracy as well as high false alarm rates, especially when new phishing methods are introduced [3]. The blacklist based approach of identifying the phishing attacks is ineffective because registering new domains has become easy, and no complete blacklist will guarantee an up to date database [3]. Effective phishing mitigation measures are required to detect active breaches and minimise the damage that successful attacks can cause. The study aims to develop a real-time browser extension empowered by machine learning algorithms aimed at identifying and classifying phishing attacks on web pages. Section II provides a literature review, highlighting previous research on phishing detection methods, their strengths, and their limitations, particularly regarding zero-day attacks. Section III details the research methodology, outlining the design, development, and implementation of the machine learning model and browser extension. Section IV presents the implementation process, including data collection, model training, and the development of the browser extension. Section V discusses the results and analysis, emphasizing the model's performance in real-time phishing detection and zero-day attack identification. Finally, Section VI concludes the study and suggests future work for enhancing the phishing detection system.

## II. Literature Review

In the paper by [4], the author introduced a phishing detection method that merges wrapper features techniques with different classification algorithms in machine learning . The classifiers used include Random Forest, Naïve Bayes, Support Vector Machine and k-Nearest Neighbor,. Through this approach, Ali succeeded to enhance the accuracy of the model's feature selection.  By using the Random Forest classifier the research achieved a 97.3% True Positive Rate and a 97% True Negative Rate in detecting the phishing websites. However, one important factor that the authors failed to account for in their methodology is the identification of zero-day attacks. Also, the wrapper feature selection method adds extra computational overhead, resulting in increased processing time. Similarly, [5] created a desktop application with the Java programming language. They presented a phishing detection system based on URL and CSS matching methodologies, assuming that phishing attempts follow the same CSS style as legitimate websites. Their proposed approach achieved a True Positive Rate of 93.27% and a True Negative Rate of 100%. Notably, it demonstrated a low to zero False Positive Rate and had the capability to handle a wide range of websites. But, the system had drawbacks, including high memory consumption and a failure to perform well when dealing with CSS files containing a large number of rules. However, the identification of zero day attacks is an element that the authors' failed to take into consideration. Similarly, [6] introduced a machine learning technique focused on examining specific elements within web addresses. Despite analyzing various aspects of URLs, the researchers utilized feature selection to narrow down to nine crucial features, chosen through multiple algorithms for their significance. This streamlined feature approach renders the model suitable for devices with limited computational capabilities. In tests conducted on the ISCXURL-2016 dataset, Random Forest (RF) emerged with the highest accuracy, reaching 96.57%. Similarly, [7] focused their study on evaluating how well machine learning approaches work when combined to counteract phishing attempts. They tested various boosting methods using various classifiers as part of their investigation. They discovered that the most promising results in differentiating between phishing and legitimate websites came from the combination of AdaBoost and SVM classifier. Through testing, they discovered that using AdaBoost with SVM produced the best accuracy in identifying phishing sites—97.61% with an f-measure of 0.976—beating out other classification techniques. Even though their method produced excellent detection accuracy, the model would improve and perform better if adding a feature selection algorithm. Another study by [8] proposes a framework which aims to simplify comparative analysis among systems with varying feature sets, addressing the dynamic nature of phishing sites and adapting to evolving tactics. The study creates a dataset in accordance with the suggested framework by means of fine-tuning classification and analysis of 87 commonly recognised traits. The experiments, employing a conceptual replication approach, assess classifier performance, feature combinations, and selection methods. Notably, Random Forest emerges as the most effective classifier. The study accuracy score was 96.61%. Filter-based feature selection methods outshine wrapper methods, boosting performance up to 96.83%. However, the authors did not consider the detection of zero-day attacks in their study. Similarly, the paper [9] introduces an advanced phishing detection model focusing on HTML content. The approach combines Multilayer Perceptron (MLP) which is good for handling organized data in tables along with two pre trained Natural Language Processing (NLP) models to understand text features better. When these models work together, they do really well at predicting phishing websites which is even better than other methods tested on a popular dataset called CatchPhish HTML. However, the paper does not explicitly discuss whether the proposed model addresses zero-day attacks. Similarly another paper by [10] reviews various machine learning techniques employed for phishing URL detection. They analyze the effectiveness of each method in detecting phishing URLs. The researchers evaluated how well each approach identified phishing URLs by comparing their accuracy, precision, recall, and F1-score. These techniques includes the decision trees, random forests, support vector machines, and deep learning models. Furthermore, their research emphasizes the significance of feature engineering in improving the effectiveness of machine learning models for detecting phishing URLs. However, the authors did not consider the detection of zero-day attacks in their study.

Traditional phishing detection methods such as blacklists and heuristics have proven that they don't work well at identifying and preventing phishing attacks especially in terms of real time detection. While machine learning algorithms have helped in improving phishing detection accuracy, there is still a notable research gap in developing techniques that can successfully identify and respond to zero day phishing attacks in real time. This gap limits the goal of real-time detection and prevention of zero-day threats. Current systems often operate offline or with a delay, which affects their ability to effectively detect phishing attacks whenever they happen. This delay makes it really hard in mitigating the impact of phishing attacks especially zero-day attacks. These attacks exploit vulnerabilities of the system even before they are identified and addressed. Hence, there is a need for research to focus on developing techniques that can swiftly identify and respond to phishing attacks in real-time and thus detecting zero-day attacks. This study will address these critical gaps in the existing literature and contribute towards the development of more effective and real-time phishing detection methods using machine learning.

III. RESEARCH METHODOLOGY

The methodology section provides a comprehensive overview of the approach taken to develop the real-time browser extension empowered by machine learning algorithms for identifying and classifying phishing attacks.

A. Research Design

The research design is outlined in Figure 1 below illustrating the high-level architectural diagram designed for the research. Phase 1 involves planning, where the aim, research question, and objectives are defined. The next step is Phase 2, which is the literature review. During this phase, the author extensively reviewed existing research on traditional phishing detection methods, machine learning for phishing, and current systems. By doing this, the author identified a gap in the research. In phase 3, i.e. the Design phase, the author selects the design for developing both the user interface (UI) and the backend. In this case, the UI refers to the browser extension, while the backend comprises the machine learning model for phishing detection. These components are then integrated together. In Phase 4, the author decides to use publicly available datasets from sources like PhishTank and Kaggle for creating the dataset. Features are extracted from these datasets. Phase 5 involves the Implementation, where Python is used for developing the machine learning model, while JavaScript, HTML, and CSS are utilized for developing the browser extension. In Phase 6, the Result Analysis is conducted

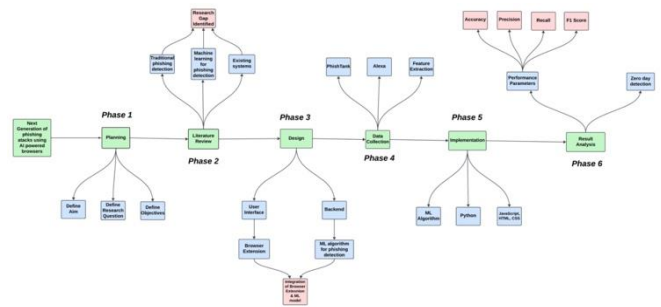

Fig. 1. High Level Design

Performance parameters such as accuracy, precision, recall, and F1 score are considered. To evaluate the model's zero-day detection capability for phishing URLs, the author chooses a 15-day testing period. Each day, 15 URLs are obtained from PhishTank reported on the same day and used for prediction with the machine learning model.

B. Data Collection & Data Analysis

Research papers, conference proceedings, and existing literature on AI-powered browser extensions and phishing detection methods are collected for a thorough understanding of the current state of the field. The author has identified key terms relevant to the research topic and had used them alongside Boolean operators to conduct a more targeted search for relevant literature [12].
The Keywords used for search are as follows:
Phishing OR Machine learning OR Real time detection OR Phishing detection techniques OR Supervised Learning OR Browser Security.

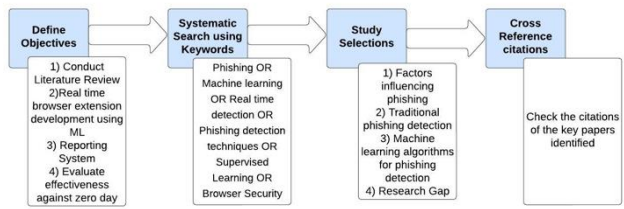

Fig. 2. Search Process

Quantitative data analysis is employed in this research to assess the effectiveness of the browser extension in real-time phishing detection, particularly against zero-day attacks. The author uses numbers and statistics to understand how well the browser extension works in detecting phishing websites especially zero day attacks. These metrics provide objective and numerical measures of the performance of the browser extension [13].

## IV. IMPLEMENTATION

### A. Machine Learning Model

The implementation of the model loads a dataset file named 'Phishing_Dataset.csv' containing labelled URLs (phishing (1) or legitimate (0)) using pandas. It splits the dataset into features (X) and the target variable (y). The dataset is further split into training and testing sets using train_test_split from sklearn.model_selection. In the code, test_size=0.3 means that 30% of the dataset will be used for testing and the remaining 70% will be used for training. The Random Forest classifier is initialized with 100 estimators and trained on the training data. This is a classifier from the scikit-learn library that implements a random forest algorithm for classification. The trained model is saved to a file called 'random_forest_model.pkl' using joblib.dump() function. Figure 3 below shows the distribution of legitimate URL and phishing URL in the acquired dataset after data pre-processing. The final dataset contains total 380,009 records with 196,757 instances classified as legitimate URLs and 183,252 instances classified as phishing URLs. Phishing URL is represented by 1 and Legitimate URL is represented by 0. Feature Extraction involved extracting specific characteristics from the dataset. The collected data is represented as features for training the machine learning model. There are 42 features considered for the dataset.

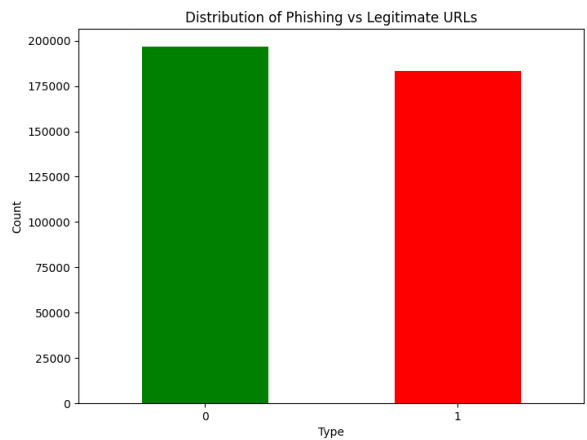

Fig. 3. Distribution of Legitimate URL & Phishing URL

### B. Development of Browser Extension

The communication between extension and ML model is done via HTTP requests. The browser extension integrates with an external API to determine the whether the entered URL is legitimate or phishing. The API endpoint receives the URL as input and returns a response indicating whether the URL is legitimate or a phishing website. If the URL is phishing then the API response will return as 'Phishing' and if the URL is legitimate the API response will return as 'Legitimate'. The first step is to host the ML model locally so that the ML model is accessible via an API endpoint. This will be done using a web framework called Flask. The author created a python script that uses Flask to create a web service for predicting whether a given URL is phishing or legitimate using the pre-trained machine learning model.

The author had also implemented a reporting system within the browser extension. This system enables users to contribute to the identification and mitigation of phishing threats directly. Users can easily flag suspicious websites whenever they encounter during browsing sessions.

## V. RESULTS & DISCUSSIONS

In this section the author focuses on the analysis and evaluation of the results obtained from the implementation of the browser extension integrated with ML model. This section discusses the zero day detection capability of the browser extension. Before enabling the browser extension, the author accessed the phishing URL https://mywkueduu.weebly.com/ obtained from PhishTank. This URL was not blocked by Google Safe Browsing at the time of testing. Figure 4 below shows the screenshot of accessing the zero day phishing website before enabling the browser extension. Upon accessing the URL, the author observed the following behaviour:
  a. The browser successfully navigated to the phishing website hosted at https://mywkueduu.weebly.com/.
  b. Despite being a known phishing URL, Google Safe Browsing did not display any warning or block the access to the website. Users who visit the website may be at risk of falling victim to phishing attacks [15].

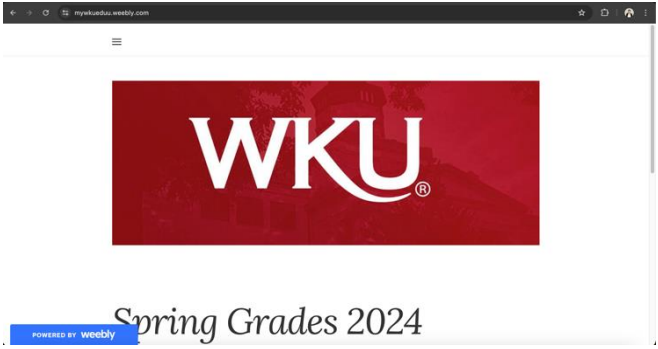

Fig. 4. Without Browser Extension

After enabling the browser extension developed for phishing website detection, the author accessed the same phishing URL https://mywkueduu.weebly.com/. The browser extension was configured to make real-time API calls to determine the legitimacy of the entered URLs. Figure 5 below shows the screenshot of accessing the zero day phishing website after enabling the browser extension. Upon accessing the URL, the author observed the following behavior:
    a. The browser extension detected that the website was a phishing site based on the response from the API.
    b. As per the extension's behavior, users were redirected to a blocked page upon accessing the phishing website.
    c. The blocked page displayed a prominent message indicating that phishing had been detected, alerting users to the potential risk.

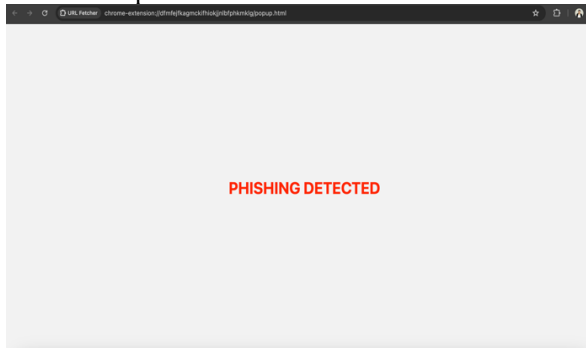

Fig. 5. Browser Extension

The results showed that the model had an accuracy of 98.32%, precision of 98.62%, recall of 97.86%, and an F1-score of 98.24%.

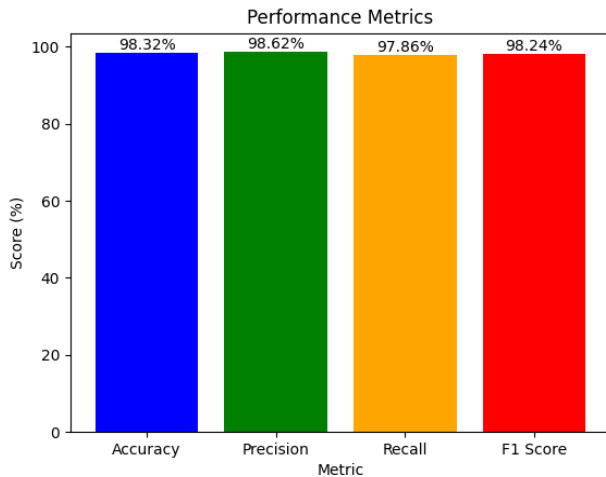

Fig. 6. Performance Metrics of Random Forest

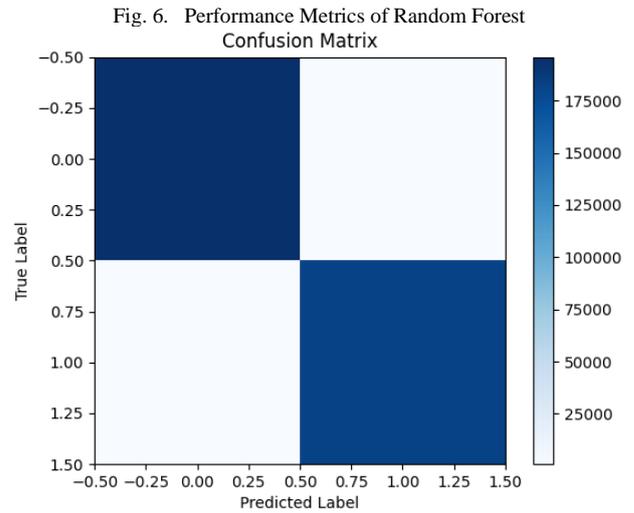

Fig. 7. Confusion Matrix of Random Forest

From the Figure 7, the analysis reveals the following:
    a. 195,879 instances correctly classified as positive (true positive cases) i.e. correctly identifying phishing URLs as phishing.
    b. 181,958 instances correctly classified as negative (true negative cases) i.e. correctly identifying legitimate URLs as legitimate.
    c. 878 instances incorrectly classified as positive when they are actually negative (false positive cases) i.e. misclassified legitimate URLs as phishing.
    d. 1,294 instances incorrectly classified as negative when they are actually positive (false negative cases) i.e. misclassified phishing URLs as legitimate.

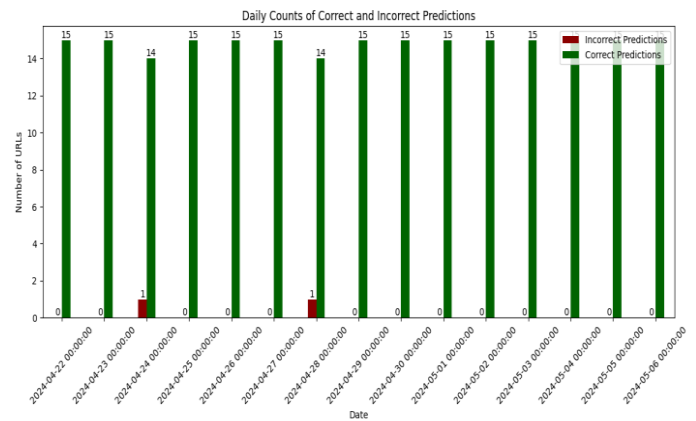

Fig. 8. Daily count of correct and incorrect predictions

The author evaluated the zero-day detection capability of the ML model for phishing URLs over a 15 day testing period. The dataset consisted of 15 phishing URLs reported on each day from Phish Tank. Figure 8 above illustrates the daily counts of correct and incorrect predictions made by the ML model over the 15-day testing period. As depicted in the figure, the ML model demonstrated remarkable capabilities in accurately identifying phishing URLs. The model consistently achieved a high rate of correct predictions throughout the testing period. Out of the 225 URLs tested over a period of 15 days, the model accurately predicted 223 URLs. Only 2 incorrect predictions occurred, where phishing

URLs were incorrectly classified as legitimate. These errors happened due to false negatives. Figure 9 below shows the overall accuracy of the model in predicting zero day phishing URLs during the 15 day testing period which is calculated at 99.11%. The zero-day testing revealed instances where the ML model successfully detected phishing websites that were undetected by Google Safe Browsing, highlighting its capability to identify zero-day phishing attacks.

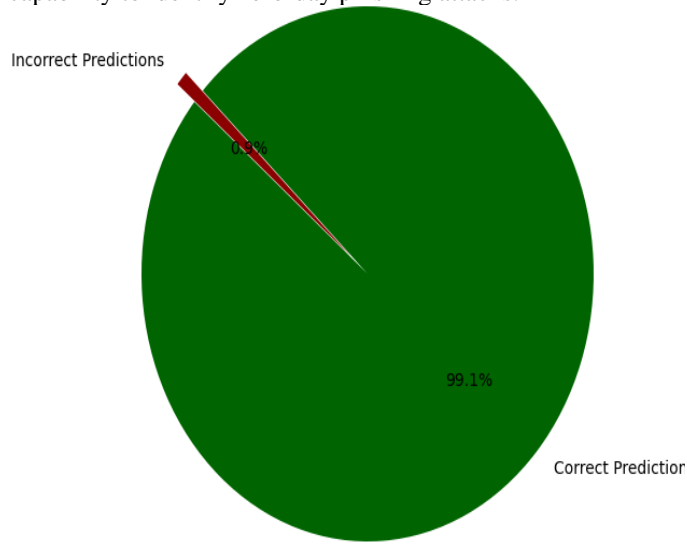

Fig. 9. Percentage of Correct and Incorrect Predictions

Figure 10 below shows the comparison of performance of Random Forest, Support vector machine, Naïve Baiyes, Decision Tree, XGboost, K Nearest Neighbor. Random Forest emerges as the top performer among these algorithms.

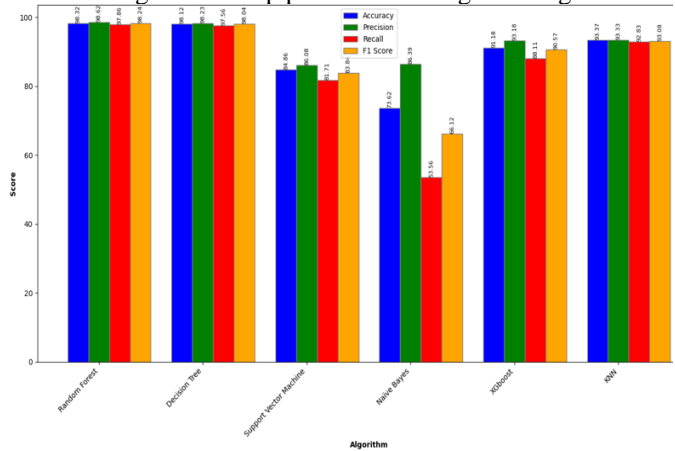

Fig. 10. Comparison of performance of different Algorithms

## VI. CONCLUSION & FUTURE WORKS

The study thoroughly explored developing and using a real-time browser extension with machine learning that can detect phishing websites. After conducting thorough experiments and analysis, the author identified several notable findings that highlight the effectiveness of this innovative approach for improving online security. The author successfully identified research gaps through an extensive literature review. The primary aim was to develop a browser extension integrated with machine learning algorithms for detecting phishing URLs. This ensured that the researcher achieved understanding of the relevant literature and advancements in the domain. The results showed that the model had an accuracy of 98.32%, precision of 98.62%, recall of 97.86%, and an F1-score of 98.24%. This achievement aligns with the objective to develop an effective solution for detecting phishing threats in real-time. The implementation of a reporting system enables users to flag suspicious websites. These reported websites are blocked by the browser extension. This achievement aligns with the objective of implementing additional features to enhance the threat mitigation capabilities of the system. The zero-day phishing attack detection testing revealed the model's remarkable capability to identify previously unseen phishing attempts. Over a 15-day testing period, the model achieved an overall accuracy rate of 99.11%, highlighting its effectiveness in detecting zero-day phishing URLs. This achievement aligns with the objective of evaluating the effectiveness against zero-day attacks. The author also observed the comparative analysis showcased the superiority of the Random Forest algorithm over other classification models, including Support Vector Machine, Naïve Bayes, Decision Tree, XGBoost, and K Nearest Neighbor. While testing the model's zero-day detection capabilities, the model performed better than traditional security measures like Google Safe Browsing. It successfully identified phishing URLs that weren't caught by these conventional methods.

For future improvements, the author suggests implementing Dynamic Dataset updates. This means automatically updating the dataset using the model's predictions. By implementing this approach, the model remains up to date on emerging phishing trends and changes. This means the model can easily detect new phishing websites whenever they are created. This will make it more effective at identifying phishing attacks in real-time. Also, integrating advanced algorithms and feature engineering with the model can enhance its accuracy and capability to detect phishing websites. By continuously refining the model with the latest techniques, we can improve its performance and adaptability to evolving threats. Finally, we can aim to integrate this solution directly into web browsers as a built-in feature. This would offer users seamless protection while browsing the internet, enhancing overall cyber security and safeguarding against phishing attacks.